\documentclass[mathleft
]{an}
\usepackage{graphicx}
\usepackage{times}
\begin{document}

% The following seven commands are intended for editorial usage and should be ignored by
% the author(s).
\Pagespan{789}{}% Document's page range. 
% If second parameter is left empty, the last page is computed automatically.
\Yearpublication{2006}%
\Yearsubmission{2005}%
\Month{11}%   
\Volume{999}%  
\Issue{88}% 
% \DOI{This.is/not.aDOI}% 

\title{Molecular Gas at High Redshift}

\author{Fabian Walter\inst{1}, Chris Carilli\inst{2} \& Emanuele Daddi\inst{3}}
%Example 
%for footnote, note the usage of the \texttt{fnmsep}
%command as separator between institute number and footnote mark} 

\titlerunning{Molecular Gas at High Redshift}
\authorrunning{Walter, Carilli \& Daddi}
\institute{
Max Planck Institut f\"ur Astronomie, Auf dem H\"ugel 71, 69117 Heidelberg, Germany
\and
National Radio Astronomy Observartory, Socorro, NM 87801, USA
\and
Laboratoire AIM, CEA Saclay,  91191 Gif-sur-Yvette Cedex, France
}

\received{30 May 2005}
\accepted{11 Nov 2005}
\publonline{later}


\abstract{% 
In order to understand galaxy evolution through cosmic
times it is critical to derive the properties of the molecular gas
content of galaxies , i.e. the material out of which stars ultimately
form. The last decade has seen rapid progress in this area, with the
detection of massive molecular gas reservoirs at high redshifts in
submillimeter--selected galaxies and quasars. In the latter case,
molecular gas reservoirs have been quantified out to redshifts z$>$6,
i.e. towards the end of cosmic reionization when the universe was less
than one Gyr old. The recent discovery of molecular gas in more normal
galaxies have extended these studies from the most extreme objects in
the unviverse (SFR$\sim$1000\,M$_\odot$\,yr$^{-1}$; quasars and
submillimeter galaxies) to more `normal' starforming systems at
redshifts 1.5--2.5 (with SFR$\sim$100\,M$_\odot$\,yr$^{-1}$). However,
detecting the molecular gas reservoirs of high--redshift galaxies that
only have moderate star formation rates
($\sim$10\,M$_\odot$\,yr$^{-1}$, similar to the faint galaxies seen in
the Hubble Ultra Deep Field) will likely have to await the completion
of ALMA.  } \maketitle

\section{Recent Progress in understanding Galaxy Formation}

The last decade has seen dramatic advances in our understanding of
galaxy formation. Cosmic geometry, the mass-energy content of the
Universe, and the initial density fluctuation spectrum, have been
constrained to high accuracy (e.g., Spergel et al.\ 2007). Given the
underlying distribution of dark matter, the main challenge of galaxy
formation studies is to understand how the baryons assemble into the
observed Universe. In this context, structure formation through
gravitational instabilities has been calculated in detail through
numerical studies, and observationally verified through studies of
galaxy distributions (e.g., Springel et al.\ 2006). The history of
star formation, and the build up of stellar mass, as a function of
galaxy type and mass, are now constrained out to the tail-end of
cosmic reionization, within 1~Gyr of the Big Bang, and pushing toward
first light in the Universe (e.g., Bouwens et al.\ 2010).

The principle results of these studies can be summarized as follows.
The comoving cosmic star formation rate density increases by more than
an order of magnitude from $z \sim 0$ to 1, peaks around $z \sim 2$ to
3, and likely drops gradually out to $z \sim 8$ (e.g., Hopkins \&
Beacom 2006, Bouwens et al.\ 2010). The build-up of stellar mass
follows this evolution, as does the temporal integral (for z$<$2,
e.g. Ilbert et al.\ 2010). The redshift range $z \sim 1.5$ to 3 has
been called the `epoch of galaxy assembly', when about half the stars
in the Universe form in spheroidal galaxies. It has been shown that
star formation shifts systematically from lower luminosity galaxies
(L$_{FIR} \sim 10^{10}$ L$_\odot$) at low redshift, to high star
formation rate galaxies (L$_{FIR} \ge 10^{11}$ L$_\odot$) at $z \sim
2$ (`downsizing', e.g., Le Floc'h et al.\ 2005, Smol{\v c}i{\'c} et
al.\ 2009). Another key finding is the tight relation between the star
formation rate and the stellar mass for star forming galaxies up to
redshifts z$\sim$3 (e.g. Noeske et al.\ 2007; Elbaz et al.\ 2007;
Daddi et al.\ 2007; Magdis et al.\ 2010). High--redshift starforming
galaxies are typically selected through optical/IR studies (e.g.,
Steidel et al.\ 1999, Daddi et al.\ 2004, van Dokkum et al.\ 2006)

Besides the star forming galaxies, a population of passively evolving,
relatively massive (stellar masses $\sim 10^{10}$ to 10$^{11}$
M$_{\odot}$) galaxies (`red and dead'), have been detected at $z > 1$,
comprising roughly 50\% of the galaxies selected in near-IR surveys
(see, e.g., review by Renzini 2006). These galaxies must form the
majority of their stars quickly at even earlier epochs (e.g., Wiklind
et al.\ 2008).

Lastly, substantial populations of supermassive black holes and
galaxies are now being detected routinely back to first light and
cosmic reionization ($z \sim 6.5$). Populations include normal star
forming galaxies, such as the Ly-$\alpha$ selected galaxies, with star
formation rates of order 10\,M$_\odot$ year$^{-1}$ (Taniguchi et al.\
2005, Ouchi et al.\ 2009) as well as quasar host galaxies, often with
star formation rates that are 100 times higher (e.g., Wang et al
2008). The latter likely represent a major star formation epoch for
very massive galaxies within 1 Gyr of the Big Bang.  Most recently,
GRBs are showing great potential to probe galaxy formation out to $z >
8$ (Tanvir et al.\ 2009, Salvaterra et al.\ 2009).

These observations have led to a new model for galaxy formation.  As
opposed to either cooling of virialized, hot halo gas, or major,
gas-rich mergers, the dominant mode of star formation during the epoch
of galaxy assembly may be driven by cold mode accretion, or stream-fed
galaxy formation (e.g., Kere{\v s} et al.\ 2005, Dekel et al.\
2009). Simulations suggest that gas accretion in early massive
galaxies occurs along cold streams from the filamentary intergalactic
medium (IGM) that never shock-heat, but streams onto the galaxy at
close to the free-fall time. This cool gas forms a thick, turbulent
rotating disk which very efficiently forms stars in a few giant
regions in the disk (e.g., Bournaud et al.\ 2007).  The star forming
regions then migrate to the galaxy center via dynamical friction and
viscosity, forming compact stellar bulges. The process leads to
relatively steady, active ($\sim 100$ M$_\odot$ year$^{-1}$) star
formation in galaxies over timescales of order 1 Gyr.  Subsequent dry
mergers at $z < 2$ lead to continued total mass build up, and
morphological evolution, but little subsequent star formation (e.g.,
Skelton et al.\ 2009, Van der Wel et al.\ 2009, Robaina et al.\ 2010).

%The %majority of stars in progenitor spheroidal galaxies may have
%been %formed via cold mode accretion at $z \sim 2$ to 3. {\bf more
%references needed}

%At first sight, the cold mode accretion model appears
%anti-hierarchical, in that massive galaxies form most of their stars
%early in the Universe, and the more massive, the earlier.  However, as
%pointed by Dave et al.\ (2007) and Crane et al.\ 2009 {\bf which
%references were you referring to here?}, the model is consistent with
%hierarchical structure formation models when density biasing is also
%included, ie. galaxy formation starts earliest in the densest regions
%of the Universe.

\section  {Molecular gas: the key to testing galaxy formation models}

While progress has been impressive, the model above is based almost
exclusively on optical and near-IR studies of the stars, star
formation, and ionized gas. There remains a major gap in our
understanding of galaxy formation, namely, observations of the cool
gas, the fuel for star formation in galaxies. In essence, current
studies probe the products of the process of galaxy formation, but
miss the source. Numerous observational and theoretical papers have
pointed out the crucial need for observations of the cool molecular
gas feeding star formation in galaxies (e.g., Dressler et al.\ 2009;
Obreschkow \& Rawlings 2009).

{\bf The molecular gas density history of the Universe:} The key to
future studies of galaxy formation is to determine the evolution of
the molecular gas density of the Universe.  Over the last two decades,
the star formation history of the Universe (SFHU) plot has been
perhaps the most fundamental tool for studying galaxy formation.  At
the same time, very detailed studies of star formation in nearby
galaxies have reached two critical conclusions.  First, star formation
relates closely with the molecular gas content, but has little
relation to the atomic neutral gas content (e.g., Bigiel et al.\
2009).  This fact is verified at high $z$ by the lack of evolution of
the cosmic HI mass density, as determined from damped Ly$\alpha$
systems, over the same redshift range where the star formation rate
density is increasing by more than an order of magnitude (Prochaska \&
Wolfe 2009). Second, once molecular gas forms, it forms stars, to
first order, according to a star formation law (e.g., Kennicutt 1998,
Bigiel et al.\ 2009, Leroy et al.\ 2009, Krumholz et al.\ 2009, Daddi
et al.\ 2010b, Genzel et al.\ 2010). In low--density environemnts, the
general idea is that, once a giant molecular cloud becomes
self-gravitating, star formation proceeds via local processes inherent
to the GMC.  If the universal star formation law were to hold to high
redshift, then the classic SFHU plot would  just be a
reflection of the molecular gas density history of the Universe.

{\bf The stellar to gas mass density ratio:} A related, and similarly
fundamental issue is to compare the stellar and gas mass of galaxies
versus redshift. Two extremes scenarios can be envisioned. At one
extreme, the gas builds up to large values (approaching 10$^{11}$
M$_\odot$), and then a dramatic starburst is triggered via, e.g., a
major gas rich merger, when the gas is rapidly converted into stars on
dynamical timescales $\sim 10^8$ years.  At the other extreme, a
continuous supply of gas comes from the IGM, cooling into molecular
clouds and forming stars at a rate comparable to the free-fall rate of
gas onto the galaxy over $\sim$~Gyr timescales. This latter case
corresponds to the cold mode accretion model. This point is currently
addressed by a number of observational programs that aim at studying
star forming galaxies at z$\sim2$ (Daddi et al.\ 2010a, 2010b, Tacconi
et al.\ 2010, Genzel et al.\ 2010, as discussed in Sec. 4.3 and 4.4 below).

More generally, the study of the cool molecular gas phase in `typical'
high high--redshift galaxies (with SFR$\sim$100\,M$_\odot$\,yr$^{-1}$)
provide key insight into the process of galaxy formation in a number
of ways:

 \begin{itemize}

 \item A universal star formation law: The low order transitions of CO
 provide
   the most direct means of measuring the total molecular gas mass.  These
   measurements can then be compared to star formation rates to
   determine the evolution of the fundamental star formation laws
   relating cool gas and star
   formation. Is there a universal relationship that
   governs the efficiency of conversion of molecular gas to stars in
   galaxies of various types
   through cosmic time?  What are the fundamental physical
   parameters behind the relationship? For a first attempt to address
   this topic see Sections 4.3 and 4.4 below.

 \item How does the ratio of molecular gas to stellar mass evolve? This ratio
   is typically $< 0.1$ in the nearby Universe, even for gas rich
   galaxies
   such as the Milky Way. Observations of a few the sBzK, and BX/BM galaxies
   provide evidence that this ratio may
   increase to $ > 1$ at $z \sim 2$ (Daddi et al.\ 2010a, Tacconi et
   al.\ 2010). 

 \item Galaxy dynamics and star formation: Molecular line observations are the
   most direct means of studying galaxy dynamics. Are these systems dominated
   by rotating disks, as predicted by cold
   mode accretion, or major mergers?  What is the relative magnitude
   of ordered motion vs.\ turbulence?  What are
   the stability criteria, and how do they compare to the distribution of star
   formation?

 \item Black hole -- bulge mass relation: Establishing a galaxy's mass
 through molecular
   gas dynamics can be used to test the evolution of the black hole --
   bulge
   mass relation back to the first galaxies. Indeed, in many cases, the
   molecular line observations are the only means to get galaxy dynamics for
   galaxies with optically bright AGN (e.g., Walter et al.\ 2003,
   Shields et al.\ 2006, Riechers et al.\ 2008, 2009).  The local
   relation suggests that the
   formation of black holes and galaxies are closely
   tied, but the origin of this relationship remains a mystery (see Sec.~4.2).

 \item ISM physics and chemistry: Detailed studies of the physical
 conditions
   in the ISM of primeval galaxies can be performed using numerous
   molecular
   line transitions, including excitation studies of CO, observations of dense
   gas tracers like CN, HCN and HCO$^+$, and other key astrochemical
   tracers. Such observations determine the temperature and density of
   the
   ISM, molecular abundances and the fraction of dense gas. To date,
   only a few extreme
   high--redshift systems have been detected in these dense gas
   tracers (e.g. Vanden Bout et al.\ 2003, Riechers et al.\ 2006,
   2007a, 2007b, Garcia--Burillo et al.\ 2006)

\end{itemize}

\section {Recent progress on molecular lines observations in early
galaxies}

\subsection{Submillimeter Galaxies (SMGs)}

Surveys at submm wavelengths have revealed highly obscured, extreme
starburst galaxies (SFR $> 1000$ M$_\odot$ year$^{-1}$) at high
redshift, the so--called submillimeter galaxies (SMGs, e.g., Smail et
al.\ 2007, Blain et al.\ 2002). Subsequent studies  have
detected large reservoirs of molecular gas in these systems ($> 10^{10}$\,M$_\odot$;
e.g., Frayer et al.\ 1998, 1999, Genzel et al.\ 2003, Greve et al.\
2005, Tacconi et al.\ 2006, 2008, Knudsen et al.\ 2009, Ivison et al.\
2010). General properties of the SMGs are summarized in the review by
Solomon \& Vanden Bout (2005) and the detections to date are
summarized in Figure~1 (as discussed in Sec. 4.3).

High--resolution imaging of SMGs by Tacconi et al.\ (2006, 2008) have
shown that their molecular gas reservoirs are compact, with a median
diameter of $\sim$4\,kpc. This morphology can be explained if these
galaxies are the results of mergers, in which the molecular gas
settles in the centre of two interacting galaxies (leading to a
starburst). Recent CO imaging of one of the most distant SMGs suggests
that cold mode accretion may also play a role in powering the ongoing
star formation (Carilli et al.\ 2010).

Chapman et al.\ (2005) have shown that the median redshift of SMGs is
z$\sim$2.3 --- the existence of a substantial high--redshift (z$>$4)
tail of the submillimeter galaxy population that host molecular gas
has recently been established by a number of studies (Schinnerer et
al.\ 2009, Daddi et al.\ 2009a, 2009b, Carilli et al.\ 2010, Coppin et
al.\ 2010).

While the study of these SMGs has opened a new window on the optically
obscured Universe at high redshift, the fact remains that these
sources are rare, likely representing the formation of the most
massive galaxies in merger-driven hyper-starbursts at high redshift,
with gas depletion timescales $< 0.1$Gyr,

\subsection{Quasars}

Likewise, surveys of molecular gas in quasar host galaxies have
revealed the presence of massive reservoirs of molecular gas in these
objects ($> 10^{10}$ M$_\odot$, i.e. comparable to the submillimeter
galaxies, e.g., Solomon \& Vanden Bout 2005, Coppin et al.\
2008). Quite remarkably, such reservoirs have now been detected in a
sizable sample of quasars at redshift 6 and greater, i.e. at the end
of cosmic reionization when the Universe was less than one Gyr old
(Walter et al.\ 2003, 2004, Bertoldi et al.\ 2003, Maiolino et al.\ 2007,
Carilli et al.\ 2007, Wang et al.\ 2010). Resolved imaging of a few
quasar hosts (Walter et al.\ 2004, Riechers et al.\ 2008a, b, Riechers
et al.\ 2009) imply that the molecular gas is extended on kpc scales
around the central black hole. It remains puzzling how such amounts of
{\em cold} molecular gas, rich in C and O, can be distributed over
many kpc scales on short (few 100 Myr) timescales.

The high FIR luminosity of the quasars, in conjunction with their
massive molecular gas reservoirs imply high star formation rates in
the host galaxies of SFR $> 1000$ M$_\odot$ year$^{-1}$. This finding
is corroborated by the detection of [CII] emission in a quasar host at
z=6.42 (Maiolino et al.\ 2005, Walter et al.\ 2009). In fact,
spatially resolved measurements of the [CII] emission were able to
constrain the size of the starburst to be emerging from the central
$\sim$1.5\,kpc of the quasar host. This implies extremely high star
formation rate surface densities
($\sim$1000\,M$_\odot$\,yr$^{-1}$\,kpc${-2}$), at the maximum what is
allowed theoretically based on Edington--limited star formation of a
radiation--pressure--supported starburst (Walter et al.\ 2009a).

The central black hole masses of the quasars that are currently
detectable reach masses of many 10$^9$ M$_\odot$. If the local
relation seen between the central black hole and surrounding bulge
mass (e.g., Magorrian et al.\ 1998, Ferrarese et al.\ 2000, Gebhardt
et al.\ 2000) were to hold at these high redshifts, one would expect
bulge masses of order 10$^{12}$\,M$_\odot$. Such high masses exceed
dynamical mass estimates of a few sources by more than an order of
magnitude (Walter et al.\ 2004, Riechers et al.\ 2008a,b, 2009), which
in turn indicates that black holes built up their masses more quickly
than the stellar bulges (at least in some quasars).

\begin{figure*}
\includegraphics[width=17cm]{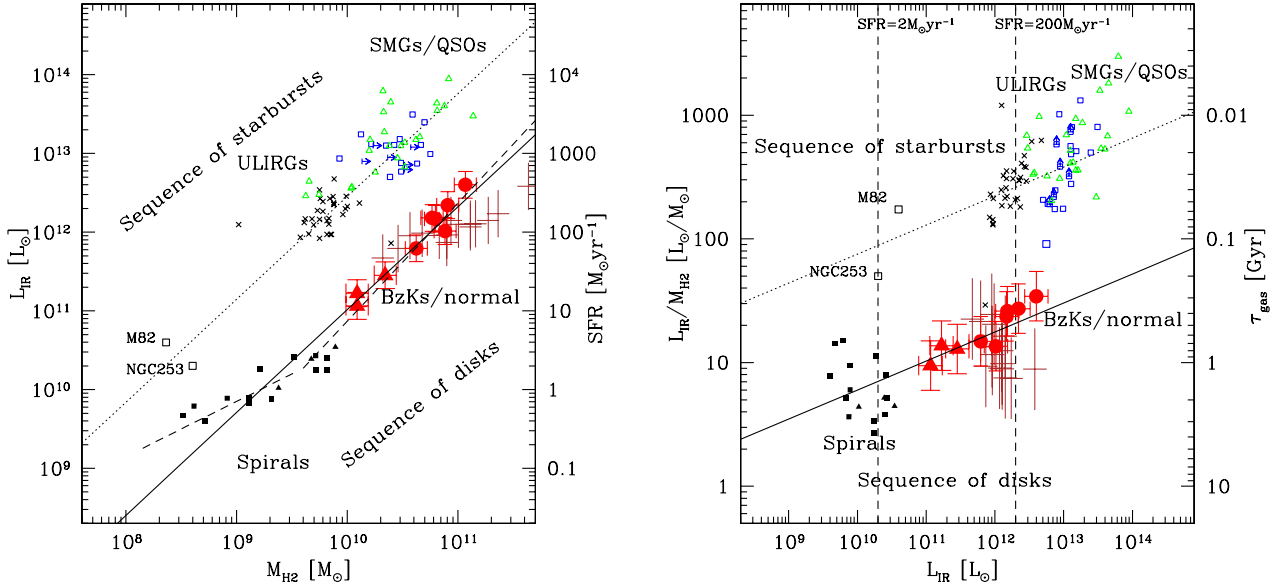}
\caption{Comparison of molecular gas masses and total
 IR bolometric luminosities (taken from Daddi et al.\ 2010b) for all high--redshifted systems detected in CO to date:
   BzK galaxies (red filled circles; Daddi et al.\ 2010a), $z\sim0.5$
   disk galaxies (red filled triangles; Salmi et al.\ in prep.),
   $z=1$--2.3 normal galaxies (Tacconi et al.\ 2010; brown crosses),
   SMGs (blue empty squares; Greve et al.\ 2005, Frayer et al.\ 2005, Daddi et al.\ 2009a, 2009b), QSOs
   (green triangles; Riechers et al.\ 2006, Solomon \& Vanden Bout
   2005), local ULIRGs (black crosses; Solomon et al.\ 1997),
      local spirals (black filled square: Leroy et al.\ 2009; black filled triangles: Wilson et al.\ 2009).
      The two nearby starbursts M82 and the nucleus of NGC~253 are
      also shown (see Daddi et al.\ 2010b for references).
   The solid line (slope of 1.31 in the left panel) is a fit
   to local spirals and BzK galaxies and the dotted line is the same relation shifted in normalization
   by 1.1~dex. 
   For guidance, two vertical lines indicate $SFR=2$ and
   200~M$_\odot$~yr$^{-1}$ in the right panel. For more details see
   Daddi et al.\ (2010b).  }  \end{figure*}

\subsection{`Normal' starforming galaxies}

More recently, major progress has been made on detecting CO emission
from more normal star forming galaxies at high $z$. These come in the form
of near-IR and color selected galaxy samples at $z \sim 1.5$ to 3.
The BzK color selection technique in particular has revealed a sample
of star forming galaxies at $z \sim 1.5$ to 2.5, selected at near-IR
wavelengths, with stellar masses in the range 10$^{10}$ to 10$^{11}$
M$_\odot$, and star formation rates $\sim 100$ M$_\odot$\,yr$^{-1}$
(Daddi et al.\ 2004). These galaxies have a space density $>10$ times
those the submm galaxies.

Daddi et al.\ (2008) showed that star forming BzK galaxies (sBzK)
uniformly contain molecular gas reservoirs of comparable mass to the
submm galaxies, and yet they are forming stars at $\sim$10 times lower
rates.  This leads to high gas depletion times of up to $\sim$1 Gyr
(see Fig.~1).  Follow--up observations showed that massive gas
reservoirs have been detected in each BzK galaxy that was targeted
(Daddi et al.\ 2010a). The implied gas fractions are very high and the
gas mass in these galaxies is comparable to or larger than the stellar
mass, and the gas accounts for 50--65\% of the baryons with the
galaxies' half--light radius. New observations of an even larger
sample of `normal' starforming galaxies have confirmed and extended
these findings (Tacconi et al.\ 2010).

In a few cases, the molecular gas emission in these objects could even
be spatially resolved (see Tacconi et al.\ 2010, Fig.~2, for a
particularly striking example) allowing for a determination of the
dynamical mass. A good constraint on the dynamical mass also allows
one to derive an independent measurement of the CO--to--H$_2$
conversion factor (as done in the BzK sample by Daddi et
al.\ 2010a). The interesting result of such an analysis is that the
BzK's have a conversion factor similar to the Galaxy, i.e. a factor of
$\sim5$ larger than what is found in local ULIRGs.

In conclusion these systems likely represent the major star formation
epoch in early massive galaxies (presumably driven by cold mode
accretion).  In the current picture these galaxies may turn into
passively evolving galaxies at $z \sim 1$, which may eventually turn
into `red and dead' cluster elliptical galaxies seen today. They are
substantially different from the quasar and submillimeter population
discussed above in terms of their larger spatial sizes, lower star
formation efficiencies, and lower molecular gas excitation
(e.g. Dannerbauer et al.\ 2009, Aravena et al.\ 2010).

\subsection{Different Star Formation Laws at High Redshift?}

The detection of molecular gas in systems spanning a wide range of
parameters now allows one to investigate the star formation law
(i.e. the dependence of the star formation rate on the available gas)
at high redshift. In Figure 1 (left), we show the summary plot of
 Daddi et al.\ 2010b. Here the infrared luminosities (a proxy for the
 SFR) is plotted as function of molecular gas mass (M$_{\rm H2}$) of
 all high--redshift objects detected in CO to date, including
 submillimeter galaxies, quasars and BzK,BM/BX galaxies (see figure
 caption for references). As discussed in Daddi et al.\ (2010b), two
 different conversion factors (to derive M$_{\rm H2}$) have been used
 for the high redhift galaxy populations in this plot: the Galactic
 one for the local spiral galaxies and the BzK/normal galaxies at
 z$\sim$1.5--2.5 and a lower `ULIRG' conversion factor for the ULIRS,
 submillimeter galaxies and quasars. Plotted this way, there are two
 sequences emerging (each of which can be fitted with a power law of
 slope $\sim$1.3), one for the `starbursts' (ULRIGS, SMGs and QSOs)
 and one for the `disk' galaxies. It should be stressed that the
 offset in sequences seen here is not only due to the choice of
 conversion factors (e.g., Fig. 13 in Daddi et al.\ 2010a). The right
 hand panel shows the same data, but the y--axis shows the ratio of
 infrared luminosity to molecular gas mass (i.e. the the gas depletion
 time) plotted as a function of total infrared luminosity. It is
 obvious from this figure that the gas depletion times in the
 BzKs/normal galaxies are about an order of magnitude longer than in
 the systems that are presumably undergoing mergers (such as the
 ULIRGs, SMGs and QSOs; see also discussion in Genzel et al.\
 2010). These different star formation efficiencies appear to be
 regulated by the dynamical timescales of the systems (Daddi et al.\
 2010b, Genzel et al.\ 2010).

These initial molecular gas observations of the different population
of star forming galaxies during the epoch of galaxy assembly present a
remarkable opportunity to study the formation of normal galaxies at
critical early epochs, when most of their stars form, and when the
dominant baryon mass constituent was gas, rather than stars.  The
galaxies are typically extended on scales $\sim\,1"$ in the optical,
providing ideal targets for follow-up high resolution imaging of the
gas using the EVLA and ALMA.

\section{Outlook: EVLA and ALMA}

Major progress in the studies of the molecular gas content of galaxies
throughout cosmic times is currently limited by the following
technological restrictions:

\begin{itemize}

\item {\em Sensitivity:} Although remarkable progress has been made in
recent years, the sensitivities of current interferometers do only
allow to study the molecular gas content in objects that have high
star formation rates and/or high molecular gas masses (of order
10$^{10}$\,M$_\odot$, SFR$\sim$100\,M$_\odot$\,yr$^{-1}$). To push this
to more typical object at high redshift (with SFR that are an order of
magnitude lower) much higher sensitivites are required. ALMA, with its
unprecedented collecting area, will without doubt revolutionize the field of
quantifying the molecular gas content in high--z systems, e.g. through
dedicated studies of molecular line deep fields.

\item{\em Resolution:} Spatially resolved imaging of galaxies remains
an important task to derive the extent of the molecular gas emission
and to constrain the dynamical masses of the systems (which in turn
give independent estimates of the conversion factor to derive
molecular gas masses from CO observations), and to search for
observational evidence for interactions and/or cold mode
accretion. Both EVLA and ALMA will enable such studies at high spatial
resolution -- but even with these new facilities such studies will be
restricted to small samples due to the intrinsic faintness of the
sources.

\item{\em Spectral Coverage:} As galaxies are observed at higher and
higher redshifts, their molecular lines are redshifted. Since the
molecular gas excitation has proven to be different in quasars,
submillimeter galaxies and more normal galaxies (e.g., Weiss et al.\
2005, 2007a, 2007b, Dannerbauer et al.\ 2009), it is important that
the same tranisions (e.g. rotational J transitions of carbon monoxide)
are observed to compare the various galaxy populations. The EVLA will
enable detailed studies of the ground transitions of CO up to high
redshift. On the other hand ALMA will enable studies of molecular gas
tracers that are currently very difficult to observe (e.g. [CII],
[NII], [OIII] and other fine structure lines, e.g., Walter et al.\
2009b).

\end{itemize}

%\begin{figure}
%\includegraphics[width=50mm,height=28mm]{testbild.eps}
%\caption{CAPTION}
%\label{label1}
%\end{figure}

\acknowledgements It is our pleasure to acknowledge our collaborators,
in particular Frank Bertoldi, Pierre Cox, Helmut Dannerbauer, Roberto
Maiolino, Roberto Neri, Dominik Riechers, Ran Wang, and Axel Weiss.

\end{document}